\documentclass[11pt,showpacs,preprintnumbers,amsmath,amssymb,prd,nofootinbib,superscriptaddress]{revtex4-2}

\usepackage{dcolumn}
\usepackage{bm}
\usepackage{ifpdf}
\usepackage{hyperref}
\usepackage{xcolor,color,graphicx,graphics,physics}
\usepackage[spanish,english]{babel}
\usepackage[latin1]{inputenc}
\usepackage[OT1]{fontenc}
\usepackage{latexsym,amssymb,amsmath,amsfonts, slashed,cancel, simpler-wick}
\usepackage{makeidx}
\usepackage{epsfig,subcaption}
\usepackage{natbib}
\usepackage{epstopdf}
\usepackage{mathrsfs}
\usepackage{hyperref}
\hypersetup{colorlinks=true, linkcolor=blue, citecolor=blue, urlcolor=blue}
\usepackage{enumerate}
\usepackage{tikz}
\usepackage{feynmp-auto}  
\usepackage{tikz-feynman}
\tikzfeynmanset{compat=1.1.0}
\usepackage{fixmath}

\everymath{\displaystyle}

\newcommand{\bea}{\begin{eqnarray}}
\newcommand{\eea}{\end{eqnarray}}

\newcommand{\orcid}[1]{\href{https://orcid.org/#1}{\includegraphics[width=10pt]{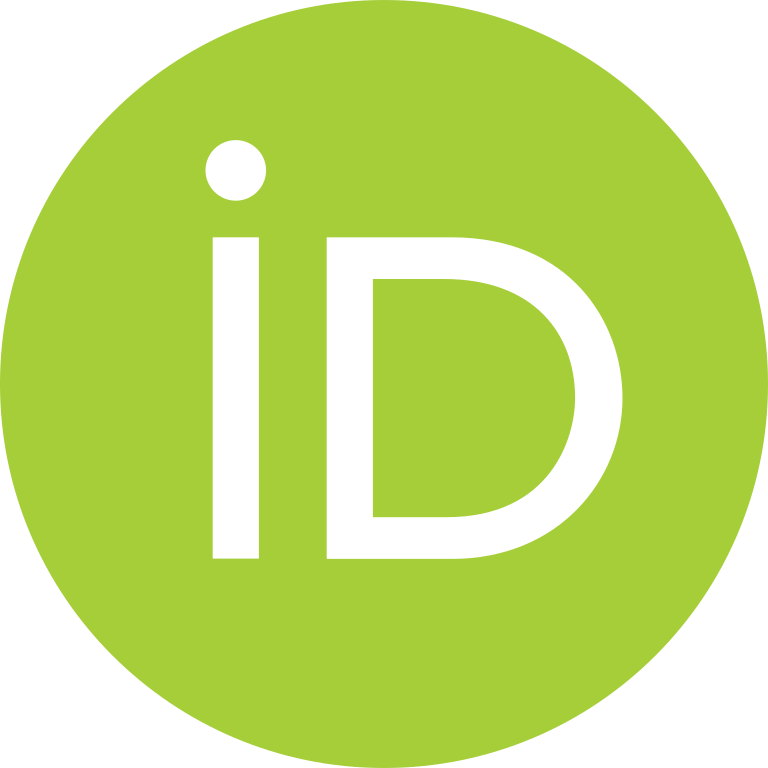}}}

\begin{document}

\title{Coulomb Potential in Podolsky-Carroll-Field-Jackiw Electrodynamics}

\author{D. S. Cabral  \orcid{0000-0002-7086-5582}}
\email{danielcabral@fisica.ufmt.br}
\affiliation{Programa de P\'{o}s-Gradua\c{c}\~{a}o em F\'{\i}sica, Instituto de F\'{\i}sica,\\ 
Universidade Federal de Mato Grosso, Cuiab\'{a}, Brasil}

\author{L. A. S. Evangelista \orcid{0009-0002-3136-2234}}
\email{lucassouza@fisica.ufmt.br}
\affiliation{Programa de P\'{o}s-Gradua\c{c}\~{a}o em F\'{\i}sica, Instituto de F\'{\i}sica,\\ 
Universidade Federal de Mato Grosso, Cuiab\'{a}, Brasil}

\author{A. F. Santos \orcid{0000-0002-2505-5273}}
\email{alesandroferreira@fisica.ufmt.br}
\affiliation{Programa de P\'{o}s-Gradua\c{c}\~{a}o em F\'{\i}sica, Instituto de F\'{\i}sica,\\ 
Universidade Federal de Mato Grosso, Cuiab\'{a}, Brasil}

\begin{abstract}

{Podolsky electrodynamics, a higher-derivative extension of Maxwell's theory characterized by the Podolsky parameter $\lambda=1/m$, which modifies the photon dispersion relation and regularizes short-distance divergences, is investigated. This framework is then coupled to the Carroll-Field-Jackiw (CFJ) model, in which a Lorentz-violating background four-vector is introduced}. Within this extended electrodynamics, the photon propagator is {obtained in the combined Podolsky-CFJ framework} and subsequently applied to M\o ller scattering. It is shown that the CFJ contribution { can reintroduce} the short-distance divergence suppressed by Podolsky's term. In the nonrelativistic limit, { both the spatial component--which introduces a preferred direction in space and thus breaks isotropy--and the timelike component--which directly affects the dispersion relation--contribute nontrivially to the interaction potential.}

\end{abstract}

\maketitle

\section{Introduction}

Podolsky's electrodynamics is a generalization of the well-established Maxwell theory. The latter predicts that a pointlike charge produces an electric field that scales as $1/r^2$, diverging in the limit $r \to 0$. In 1942, Podolsky proposed an extension of electrodynamics aimed at regularizing the high-frequency divergence problem \cite{podolsky1942generalized, podolsky1944generalized}. Unlike Proca's theory, in which the introduction of a photon mass explicitly breaks gauge symmetry \cite{weinberg1995quantum, peskin2018introduction}, Podolsky's formulation incorporates a higher-derivative term accompanied by a characteristic parameter, referred to as the Podolsky parameter. This modification naturally gives rise to a massive photon sector while preserving gauge invariance, thereby yielding a finite and non-zero field value at the origin \cite{podolsky1948review}. In this framework, the presence of a higher-derivative kinetic term in the Lagrangian establishes a lower bound for the photon mass, defined through the Podolsky parameter $\lambda = 1/m$. Various estimates for this bound have been proposed in the literature: approximately $4$ MeV, obtained from comparisons between cosmic microwave background data and the Stefan-Boltzmann law \cite{pod1}; about $35.51$ MeV, derived from corrections to the hydrogen atom ground state due to deviations from the Coulomb potential \cite{pod2}; and nearly $37$ GeV, inferred from analyses of the primordial universe expansion within a generalized electrodynamics framework \cite{pod3}. These estimates indicate that Podolsky-type effects could, in principle, manifest within measurable ranges. Nevertheless, in the absence of experimental evidence for Podolsky photons, such bounds remain insufficient to confirm their existence.

In light of these considerations, a convenient value for the Podolsky parameter, consistent with a still-undetectable photon mass, was adopted in a recent work, namely $\lambda = 3 \times 10^{-7},\text{GeV}^{-1}$ \cite{cabral2025electron}. Moreover, several studies have investigated Podolsky electrodynamics in diverse contexts, including scattering processes \cite{bufalo2014causal}, the Casimir effect \cite{barone2016casimir}, and modifications to the Stefan--Boltzmann law \cite{bonin2010podolsky}, among others. In the present study, attention is focused on the Coulomb potential within generalized electrodynamics, incorporating the Carroll--Field--Jackiw model, which introduces a background vector field that explicitly breaks Lorentz symmetry.

The Standard Model (SM) stands as one of the most successful frameworks in particle physics, founded fundamentally on Lorentz invariance, a cornerstone symmetry of modern physics. However, any theoretical framework that allows for Lorentz violation (LV) signals the presence of new physics beyond the SM. Studies on LV have been extensively conducted in various contexts \cite{colladay1998lorentz, kostelecky2002signals, kostelecky2004gravity, kostelecky2009electrodynamics}, with these foundations being applied to a wide range of phenomena, including scattering processes \cite{cabral2024lorentz, cabral2024exploring}, the Casimir effect \cite{correa2023aether, ferreira2022tfd}, and even within the framework of General Relativity \cite{jesus2019ricci, santos2015godel}, among others. Moreover, theories that seek to describe the quantum nature of gravity, such as String Theory and Loop Quantum Gravity \cite{kostelecky1989phenomenological, kostelecky1990nonperturbative, gambini1999nonstandard, alfaro2000ha}, predict the possibility of Lorentz and CPT violation near the Planck scale. Consequently, the investigation of LV effects may offer valuable insights into the manifestation of such phenomena, for instance, through modifications in the dispersion relations of elementary particles, particularly the photon.

In this context, the Carroll--Field--Jackiw (CFJ) model was formulated as an extension of Maxwell electrodynamics to investigate Lorentz and CPT violations in field theory \cite{carroll1990limits}. This model emerges from the generalization of the Chern--Simons Lagrangian to $(3+1)$ dimensions, introducing a background four-vector field that explicitly breaks Lorentz symmetry and parity while preserving gauge invariance. Within the framework of the Standard Model Extension (SME), the gauge sector contains both CPT-odd and CPT-even contributions, among which the CFJ term is included. These contributions are directly associated with vacuum birefringence \cite{kostelecky2002signals, kostelecky2006sensitive}, leading to modifications in light propagation, such as polarization-dependent velocities and a rotation of the polarization plane \cite{casana2008lorentz}. Such features highlight the importance of studying the CFJ model, as it provides a framework for a more comprehensive understanding of possible cosmological effects, including modifications in the radiation of the cosmic microwave background (CMB).

Motivated by these considerations, the present work aims to investigate the behavior of Podolsky's generalized electrodynamics, a higher derivative extension of Maxwell's theory that preserves gauge invariance while introducing the Podolsky parameter to regularize short-distance divergences, when coupled to the CFJ model. Particular emphasis is placed on how Lorentz-violating contributions manifest in potential calculations, specifically within the context of M\o ller scattering. As a result, a generalized photon propagator is derived that incorporates the CFJ contribution, representing a feature not found in previous studies in the literature.

This paper is organized as follows. In Section \ref{Section2}, the Lagrangian density of Podolsky electrodynamics in the presence of Lorentz-violating contributions is presented, and the corresponding photon propagator for the theory is derived. Section \ref{section3} is devoted to the analysis of M\o ller scattering, starting from the transition amplitude and its representation for both relevant Feynman diagrams. Based on this formulation, the interaction potential is calculated, with emphasis on the modifications introduced by the Podolsky parameter and the CFJ term. Finally, concluding remarks are provided in Section \ref{Section4}.

\section{The generalized photon propagator in the presence of a CFJ term}\label{Section2}

In this section, our main objective is to obtain the generalized photon propagator in the presence of a CFJ term. Here, generalized refers to the modification of standard QED by the inclusion of the Podolsky term. The Podolsky term introduces higher-order derivatives into the electromagnetic field, leading to a theory that naturally regularizes ultraviolet divergences and modifies the photon propagator at short distances. The total Lagrangian density describing this generalized QED in the presence of the CFJ term is given by
\begin{equation}
	\mathcal{L}=-\frac{1}{4}F_{\mu\nu}F^{\mu\nu}+\frac{\lambda^2}{2}\partial_\mu F^{\mu\nu}\partial_\alpha F^\alpha_\nu+\frac14\varepsilon^{\mu\nu\alpha\rho}v_\mu A_\nu F_{\alpha\rho}-\frac{1}{2\xi}\left(\partial_\mu A^\mu\right)^2+\overline{\psi}\left(i\gamma^\mu\partial_\mu-m\right)\psi-e\overline{\psi}\gamma^\mu A_\mu\psi,\label{Lagrangiana}
\end{equation}
where $F_{\mu\nu}=\partial_\mu A_\nu-\partial_\nu A_\mu$, $\lambda$ is the Podolsky parameter, $\xi$ is the gauge-fixing parameter, and $v_\mu$ denotes the Lorentz-violating background vector.  
It is well known that the CFJ contribution can be rewritten such that the equations of motion of a free photon in momentum space take the form
\begin{equation}
	\left(-k^2g_{\mu\nu}+k_\mu k_\nu+\lambda^2 g_{\mu\nu}k^4-\lambda^2 k^2 k_\mu k_\nu+i\varepsilon_{\mu\rho\alpha\nu}v^\rho k^\alpha-\frac{1}{\xi}k_\mu k_\nu\right)\epsilon^\mu(k)=0
\end{equation}
with $\epsilon^\mu(k)$ denoting the polarization vector.

Since both the Podolsky parameter and the CFJ term contribute to the kinetic sector, 
the problem to be solved in order to obtain the photon propagator is
\begin{equation}
	\left(-k^2g_{\mu\nu}+k_\mu k_\nu+\lambda^2 g_{\mu\nu}k^4-\lambda^2 k^2 k_\mu k_\nu+i\varepsilon_{\mu\rho\alpha\nu}v^\rho k^\alpha-\frac{1}{\xi}k_\mu k_\nu\right)D^{\mu\alpha}(k)=\delta^\alpha_\nu, \label{prop1}
\end{equation}
where $D^{\mu\alpha}(k)$ denotes the generalized photon propagator to be determined.

The longitudinal and transverse projectors are defined as
\begin{equation}
	\theta_{\mu\nu}=g_{\mu\nu}-\omega_{\mu\nu},\quad\quad \omega_{\mu\nu}=\frac{k_\mu k_\nu}{k^2},
\end{equation}
together with the Lorentz-violating tensor
\begin{equation}
	S_{\mu\nu}=\varepsilon_{\mu\rho\alpha\nu}v^\rho k^\alpha,
\end{equation}
so that Eq. \eqref{prop1} can be expressed as
\begin{equation}
	\left[-k^2(1-\lambda^2k^2)\theta_{\mu\nu}-\frac{k^2}{\xi}\omega_{\mu\nu}+iS_{\mu\nu}\right]D^{\mu\alpha}(k)=\delta^\alpha_\nu.
\end{equation}

If the operator inside the brackets is denoted by
\begin{equation}
	\Gamma_{\mu\nu}=a\theta_{\mu\nu}+b\omega_{\mu\nu}+cS_{\mu\nu},
\end{equation}
an equation of the type
\begin{equation}
	\Gamma_{\mu\nu}D^{\mu\alpha}(k)=\delta_\nu^\alpha
\end{equation}
is obtained, which must be solved to determine the generalized photon propagator.

A general ansatz for the solution may be written as
\begin{equation}
	D^{\mu\alpha}(k)=X_1g^{\mu\alpha}+X_2\omega^{\mu\alpha}+X_3S^{\mu\alpha}+X_4\frac{v^\mu v^\alpha}{v^2}+X_5\frac{v^\mu k^\alpha}{k\cdot v}+X_6\frac{v^\alpha k^\mu}{k\cdot v},
\end{equation}
which permits the propagator to be expressed as
\begin{align}
	D^{\mu\alpha}(k)=&\frac{1}{a^2+c^2l^2-c^2k^2v^2}\Biggl[a\theta^{\mu\alpha}+c^2\frac{(a-b)l^2-ak^2v^2}{ab}\omega^{\mu\alpha}+c S^{\mu\alpha}-\frac{c^2k^2}{a}v^{\mu}v^\alpha\nonumber\\&+\frac{c^2l}{a}\left(v^\mu k^\alpha+v^\alpha k^\mu\right)\Biggr],\label{definition}
\end{align}
where $l=k\cdot v$ has been defined. Recalling Eq. \eqref{definition}, the coefficients are chosen as $a=-k^2(1-\lambda^2k^2)$, $b=-k^2/\xi$, $c=i$, resulting in
{\small\begin{align}
	D_{\mu\nu}(k)=&-\frac{1}{k^4(1-\lambda^2k^2)^2+k^2v^2-(k\cdot v)^2}\Biggl[k^2(1-\lambda^2k^2)\theta_{\mu\nu}-\frac{\lambda^2(k\cdot v)^2+(1-\lambda^2k^2)v^2}{(1-\lambda^2k^2)}\omega_{\mu\nu}\nonumber\\&-iS_{\mu\nu}+\frac{v_{\mu}v_{\nu}}{1-\lambda^2k^2}-\frac{(k\cdot v)}{k^2(1-\lambda^2k^2)}(v_{\mu} k_\nu+v_\nu k_\mu)\Biggr],\label{propagator}
\end{align}}
where the Feynman gauge condition $\xi=1$ has been applied.  

The above expression corresponds to the photon propagator in generalized electrodynamics with a CFJ Lorentz-violating term. {It is important to emphasize that, although the tensorial structure of the propagator resembles that of the CFJ theory, the presence of the Podolsky term modifies the kinetic operator in a nontrivial manner, thereby significantly altering the dynamical content of the propagator. In particular, this denominator is not simply given by a scalar factor multiplying the CFJ contribution, but instead takes the form of a higher-order polynomial in the momentum. Moreover, this structure leads to the following dispersion relation}
\begin{equation}
	k^4(1-\lambda^2k^2)^2+k^2v^2-(k\cdot v)^2=0,
\end{equation}
which is consistent with the results reported in other works \cite{ferreira2025considerations}.

{This modified dispersion relation exhibits a richer pole structure than those found in the pure CFJ and Podolsky cases, reflecting the nontrivial interplay between higher-derivative dynamics and Lorentz-violating effects. In appropriate limits, it reproduces the standard massless mode of QED, while the additional solutions emerge from the combined contributions of the higher-derivative Podolsky sector and the Lorentz-violating CFJ term.}

In the next section, the result obtained here will be employed to calculate the Lorentz-violating Coulomb potential in this generalized electrodynamics.

\section{Lorentz-violating Coulomb potential in generalized electrodynamics}\label{section3}

\begin{figure}[!h]
	\centering
	\includegraphics[width=0.6\linewidth]{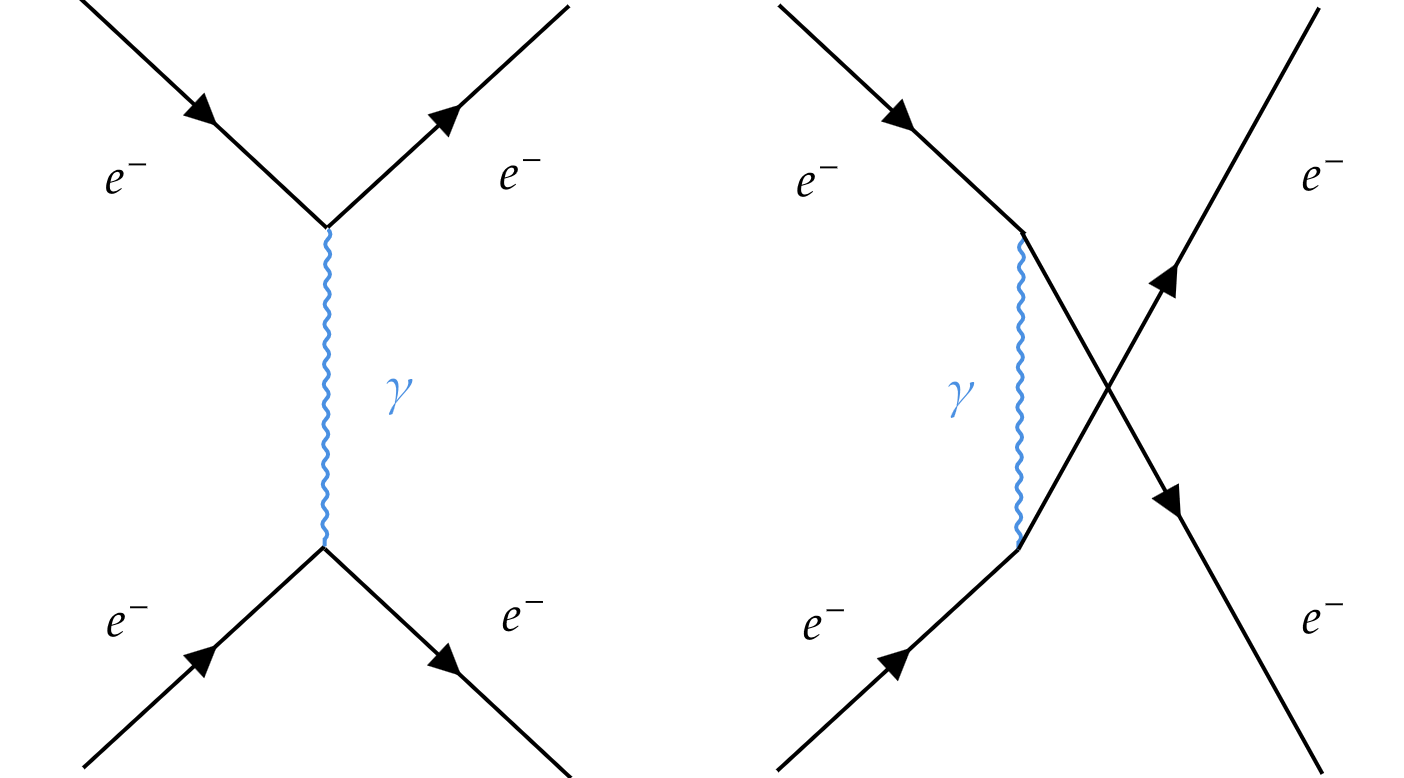}
	\caption{Feynman diagrams for M\o ller scattering. The first diagram corresponds to the $t$-channel, while the second represents to the $u$-channel. Time flows from left to right.}
	\label{diagram}
\end{figure}

This section is devoted to applying the concepts previously introduced in order to calculate the M\o ller potential. This process is represented by the Feynman diagrams shown in Figure~\ref{diagram}.

To calculate the potential associated with this process, {we'll follow the standard QED procedure for  M\o ller scattering}. With this in mind, the transition amplitude can be written as
\begin{align}
	\mathcal{M}_{fi}&=\mathcal{M}_t+\mathcal{M}_u\nonumber\\
	&=D_{\mu\nu}(q)\left[\Bar{u}(p_3)V^\mu u(p_1)\Bar{u}(p_4)V^\nu u(p_2)\right]-D_{\mu\nu}(q^\prime)\left[\Bar{u}(p_4)V^\mu u(p_1)\Bar{u}(p_3)V^\nu u(p_2)\right],
\end{align}
where $\mathcal{M}_t$ and $\mathcal{M}_u$ correspond to the $t$- and $u$-channel contributions, respectively; $V^\mu = i e \gamma^\mu$ denotes the usual QED vertex; $D_{\mu\nu}(q)$ represents the Podolsky photon propagator in the presence of the Lorentz-violating CFJ term (see Eq.~\eqref{propagator}); and the momentum transfer is given by $q = p_1 - p_3$ for the $t$-channel and $q' = p_1 - p_4$ for the $u$-channel.

The main objective is to determine the potential, a nonrelativistic quantity that characterizes the effective interaction between two fermions within the framework of generalized electrodynamics with Lorentz violation. For this purpose, only the $t$-channel diagram is considered, as the $u$-channel contribution is purely relativistic. By substituting the propagator into $\mathcal{M}_t$, the following expression is obtained
\begin{align}
	\mathcal{M}_t&=-\frac{e^2}{A}\big\{q^2 \Bar{u}(p_3)\;\slashed{v}\; u(p_1)\Bar{u}(p_4)\;\slashed{v}\; u(p_2)+q^4(\lambda^2 q^2-1)^2 \Bar{u}(p_3)\gamma_\nu u(p_1)\Bar{u}(p_4)\gamma^\nu u(p_2)\nonumber\\
	&-[\lambda^2(q\cdot v)^2+(\lambda^2 q^2-1)(\lambda^2 q^4-q^2-v^2)]\Bar{u}(p_3)\;\slashed{q}\; u(p_1)\Bar{u}(p_4)\;\slashed{q}\; u(p_2)\nonumber\\
	&+(q\cdot v)\Bar{u}(p_3)\;\slashed{q}\; u(p_1)\Bar{u}(p_4)\;\slashed{v}\; u(p_2)+(q\cdot v)\Bar{u}(p_3)\;\slashed{v}\; u(p_1)\Bar{u}(p_4)\;\slashed{q}\; u(p_2)\big\},\label{M3}
\end{align}
where, for convenience, the following definition is introduced
\begin{equation}
	A=q^2(\lambda^2 q^2-1)\left[q^4(1-\lambda^2q^2)^2+q^2v^2-(q\cdot v)^2\right].
\end{equation}

In the non-relativistic limit, the particle spinor and momentum transfer are expressed as
\begin{equation}
	u(p)=\sqrt{m}u_0;\quad\quad \text{and}\quad\quad (p^\prime-p)^2=-|\Vec{p^\prime}-\vec{p}|^2+\mathcal{O}(\vec{p}^4), 
\end{equation}
where $u_0$ is the rest-frame spinor of the particle. In this regime, it is found that $\Bar{u}(p^\prime)\gamma^\mu u(p)=2m g^{\mu0}$. In addition, the momenta are written as
\begin{equation}
	p=(m,\vec{p});\quad\quad 	k=(m,\vec{k});\quad\quad p^\prime=(m,\vec{p^\prime});\quad\quad k^\prime=(m,\vec{k^\prime}).
\end{equation}
Using all these definitions, Eq.~\eqref{M3} can be written as
\begin{align}
	\mathcal{M}_t&=-\frac{4m^2e^2q^2(\lambda^2q^2-1)}{q^4(1-\lambda^2q^2)^2+q^2v^2-(q\cdot v)^2}+\frac{4m^2e^2v_0^2}{[q^4(1-\lambda^2q^2)^2+q^2v^2-(q\cdot v)^2](\lambda^2q^2-1)}\nonumber\\&=\mathcal{M}_t^{(1)}+\mathcal{M}_t^{(2)}.
\end{align}
In this context, the potential can be expressed directly as
\begin{eqnarray}
	V(r)&=&-\frac{1}{4m^2}\int\frac{d^3q}{(2\pi)^3}\mathcal{M}_{t}^{(1)}e^{i\vec{q}\cdot\vec{r}}-\frac{1}{4m^2}\int\frac{d^3q}{(2\pi)^3}\mathcal{M}_{t}^{(2)}e^{i\vec{q}\cdot\vec{r}}\nonumber\\
    &=&V^{(1)}(r)+V^{(2)}(r),\label{eq01}
\end{eqnarray}
where, after integration, each contribution to the potential, { up to second order on the Lorentz violation}, takes the form {
{\small
\begin{align}
	V^{(1)}(r)&=\frac{e^2}{4\pi r}\left(1-e^{-r/\lambda}\right)+\frac{e^2 |\vec{v}|^2 e^{-r/\lambda}}{64 \pi r^3}
\Bigg\{
 r^4 \left[2\vartheta^2 + (5 - 8\vartheta^2)e^{r/\lambda} - 1\right]
+ 2\lambda (9\vartheta^2 - 4) r^3 \nonumber\\
& - 12 \lambda^2 r^2 \left[-4\vartheta^2 + (4\vartheta^2 - 3)e^{r/\lambda} + 1\right] - 48\lambda^4 (e^{r/\lambda} - 1)
+ 48 \lambda^3 r\nonumber\\
& + \cos(2\alpha)\Big[(12\lambda^2 + r^2 + 6\lambda r)^2  - e^{r/\lambda}(144\lambda^4 + r^4 - 12\lambda^2 r^2)\Big]
\Bigg\},\label{eq02}
\end{align}
}
and
\begin{equation}
	V^{(2)}(r)=-\frac{e^2 \vartheta^2 |\vec{v}|^2  \left[4  \left(6 \lambda ^2+r^2\right)-re^{-\frac{r}{\lambda }} (9 \lambda +r)\right]}{32
   \pi  r},\label{eq10}
\end{equation}
where,} $\alpha$ is the angle between $\vec{v}$ and $\vec{r}$, while $\vartheta^2=v_0^2/|\vec{v}|^2$ defines the Lorentz-violating parameter. The integral in Eq.~\eqref{eq01} was evaluated using a set of identities and relations provided in Appendix~\ref{append}.

\begin{figure}[ht]
   \centering
        \includegraphics[width=0.7\textwidth]{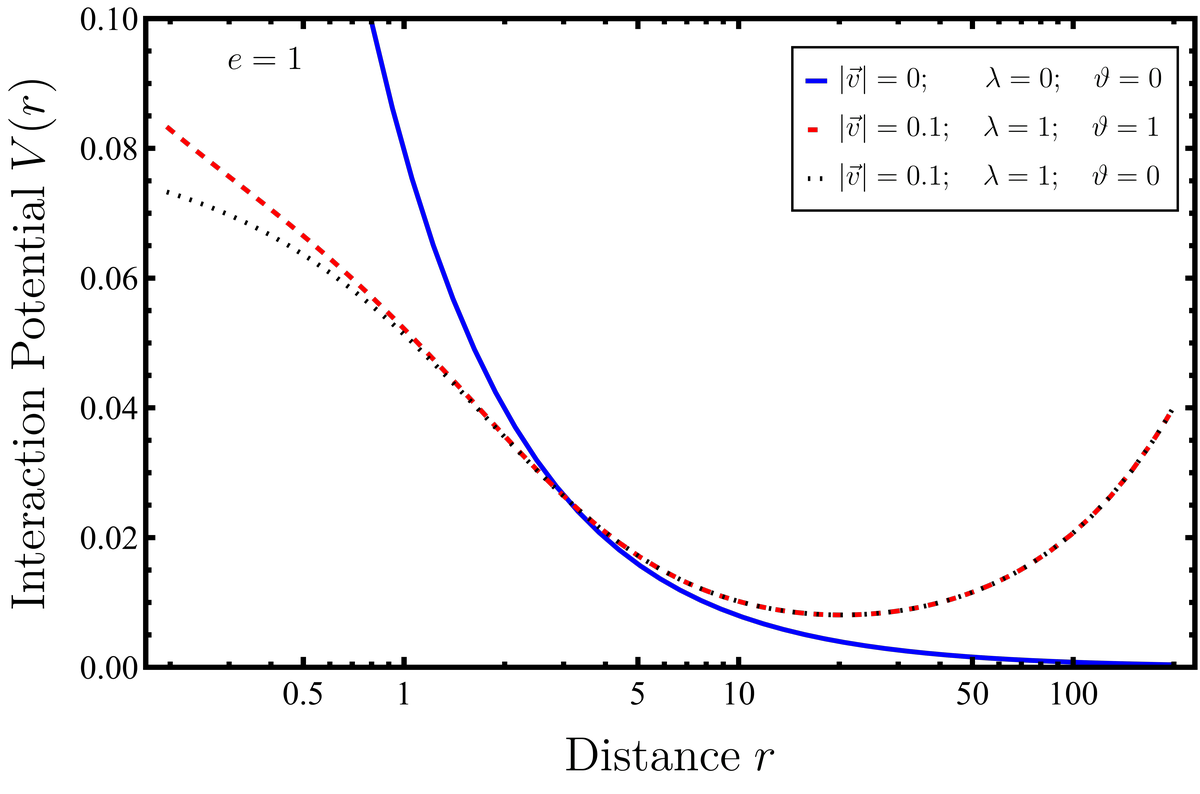}
 
    \caption{{\ Coulomb potential for different values of the parameters $\lambda$, $|\vec{v}|$ and $\vartheta$. Natural units are used, with the electric charge set to $e = 1$.\label{figtotal}}}
\end{figure}

{ For an arbitrary direction of the Lorentz-Violating field $\alpha$, the following expression is obtained
{\small\begin{align}
	V(r)&=\frac{e^2}{4\pi r}\left(1-e^{-r/\lambda}\right)+\frac{e^2 |\vec{v}|^2}{64 \pi r^3} \Biggl[
5r^4 + 36\lambda^2 r^2 - 48\lambda^4 - \cos(2\alpha)\bigl(r^4 - 12\lambda^2 r^2 + 144\lambda^4\bigr) \nonumber\\
&+ e^{-r/\lambda}\Biggl( -r^4 - 8\lambda r^3 + 12\lambda^2 r^2(4\vartheta^2-1) + 48\lambda^3 r + 48\lambda^4 +\cos(2\alpha)\,\bigl(r^2 + 6\lambda r + 12\lambda^2\bigr)^2 \Biggr) \Biggr].\label{eq30}\end{align}}}
This function is shown in Figure \ref{figtotal}.
The values of the Podolsky and CFJ parameters shown in Figure~\ref{figtotal} are purely illustrative and are used for visualization purposes only. The most reliable bound for the Podolsky parameter, consistent with a still-undetectable photon mass, is $\lambda = 3 \times 10^{-7}\,\text{GeV}^{-1}$, {Furthermore, the Carroll--Field--Jackiw vector $v_\mu$ corresponds to the $k_{AF}^\mu$ coefficient of the SME proposed in Ref.~\cite{colladay1998lorentz}, as can be seen from the Lagrangian in Eq.~\eqref{Lagrangiana}.} In that work, an upper bound of order $10^{-42}\,\text{GeV}$ was established for its time-like component, and about $10^{-41}\,\text{GeV}$ for the space-like contribution.

It is well known that the introduction of Podolsky's parameter removes the singularity of the Coulomb potential at $r = 0$ \cite{podolsky1942generalized}. To analyze the behavior of the generalized Coulomb potential within the CFJ framework, we return to Eq.~\eqref{eq02} in the short-distance limit, namely,
{\begin{equation}
	V(r)\approx\frac{3e^2|\vec{v}|^2\vartheta^2\lambda^2}{4\pi r}+\frac{(4+5|\vec{v}|^2\lambda^2-12|\vec{v}|^2\vartheta^2\lambda^2)e^2}{16\pi\lambda}.
\end{equation}}
In other words, while standard QED diverges as $V(r) \sim 1/r$ and Podolsky's theory provides a form of regularization, yielding $V(r) \sim 1/(4 \pi \lambda)$, the Lorentz-violating contribution introduces a correction to this finite term along with a new pole of the form { $\sim \vartheta/r$, which reintroduces a divergence at the origin if $\vartheta\neq0$.}
On the other hand, in the large-distance regime, the dominant term of Eq.~\eqref{eq02} becomes
{\begin{equation}
	V(r)\approx \frac{e^2|\vec{v}|^2\left[5-\cos(2\alpha)\right]}{64\pi}r,
\end{equation}}
leading to a linearly growing potential. In terms of physical estimates, assuming $|\vec{v}| \sim 10^{-41}$, the distance required for the interaction potential between two electrons to reach 1 Volt, before effectively vanishing, is approximately $r = 6.75 \times 10^{78}\,\text{m}$, or about $7 \times 10^{62}$ light-years, {considering $\alpha=0$ to the background field orientation.}

\section{Conclusion}\label{Section4}

In this work, M\o ller scattering was analyzed within the framework of generalized electrodynamics coupled to a CFJ Lorentz-violating background. As a first step, the modified photon propagator in this combined Podolsky--CFJ framework was derived, a result that, to the best of our knowledge, has not previously appeared in the literature. From this propagator, the corresponding dispersion relation was obtained, showing that the energy spectrum is explicitly influenced by both the Podolsky parameter and the Lorentz-violating contributions, with a clear distinction between the time- and space-like components of the background vector.

The $t$-channel diagram for $e^- e^- \to e^- e^-$ scattering was then considered in the non-relativistic limit, from which the generalized Lorentz-violating form of the Coulomb potential was obtained. In this formulation, { the time-like component $v_0$, encoded through the parameter $\vartheta$ and affecting the energy shift of the system, plays a very significant role by contributing directly to the interaction, in a manner similar to the spatial components of $v_\mu$. The main difference between them is that the temporal contribution primarily dictates the behavior at short distances $r \approx 0$. The total results also depend on the physical configuration, namely, the orientation of the spatial components with respect to the position vector in the rest frame of the charges. This feature is functionally described by an angular parameter $\alpha$.}

It was confirmed that in standard QED the Coulomb potential diverges at the origin and vanishes at large distances, whereas the introduction of the Podolsky parameter $\lambda$ removes the short-distance divergence while preserving the long-range behavior. When the CFJ term is included, however, the singularity at $r=0$ is restore, { if the Lorentz-violating field has non-vanish time-like component}. At large distances, the potential initially decays similarly to the Podolsky case, but at scales of $r \sim |\vec{v}|^{-2}$, the CFJ contribution dominates, leading to a linear growth and a divergence as $r \to \infty$.

\section*{Acknowledgments}

This work by A. F. S. is partially supported by National Council for Scientific and Technological
Development - CNPq project No. 312406/2023-1. D. S. C. and L. A. S. E. thank CAPES for financial support.

\section*{Data Availability Statement}

No Data associated in the manuscript.

\section*{Conflicts of Interest}

No conflict of interests in this paper.

\appendix
{\section{Some useful integrals}\label{append}
It can be noted that all integrals in Eq.~\eqref{eq01} appear divergent at first glance; however, they can be evaluated through a regularization procedure, taking into account the lowest order in the Lorentz-violating parameter $|\vec{v}|$. This can be done systematically as follows, considering
\begin{equation}
I=\int\frac{d\Omega}{(2\pi)^3}\int_0^\infty \frac{(1+\lambda^2q^2)q^2 e^{iqr\cos{\theta}}dq}{q^2(1+\lambda^2q^2)^2+|\vec{v}|^2\left(1-\vartheta^2-\cos^2\gamma\right)}\label{eq05}
\end{equation}
where $d\Omega=\sin{\theta}d\theta d\phi$ is the solid angle element, $\cos{\gamma}=\cos{\theta}\cos{\alpha}+\sin{\theta}\sin{\alpha}\cos{\phi}$. Here, $\gamma$ is the angle between $\vec{v}$ and $\vec{q}$, while $\alpha$ is the angle between $\vec{v}$ and $\vec{r}$.

To proceed with an expansion in the regime of small Lorentz-violating effects, we need to split \eqref{eq05} into two regions, $I=R_1(\epsilon)+R_2(\epsilon)$ for a small $\epsilon$, where
\begin{equation}
R_1(\epsilon)\approx \int\frac{d\Omega}{(2\pi)^3}\int_0^\epsilon \frac{q^2 e^{iqr\cos{\theta}}dq}{q^2+|\vec{v}|^2\left(1-\vartheta^2-\cos^2\gamma\right)}=-\frac{|\vec{v}|^2}{6\pi^2\epsilon}(3\vartheta^2-2)+\mathcal{O}(|\vec{v}|^3)\label{eq06}
\end{equation}
while
\begin{align}
R_2(\epsilon)&=\int\frac{d\Omega}{(2\pi)^3}\int_\epsilon^\infty \frac{(1+\lambda^2q^2)q^2 e^{iqr\cos{\theta}}}{q^2(1+\lambda^2q^2)^2+|\vec{v}|^2\left(1-\vartheta^2-\cos^2\gamma\right)}\nonumber\\&=\int\frac{d\Omega}{(2\pi)^3}\int_\epsilon^\infty \left[1-\frac{|\vec{v}|^2\left(1-\vartheta^2-\cos^2\gamma\right)}{q^2(1+\lambda^2q^2)^2}+\mathcal{O}(|\vec{v}|^2)\right] \frac{e^{iqr\cos{\theta}}dq}{1+\lambda^2q^2}.\label{eq07}
\end{align}
Defining $R_2=A+B$, one can see that the first contribution of \eqref{eq07} can be evaluated by employing distribution theory and the residue theorem. Proceeding, one obtains
\begin{equation}
	A=\frac{1}{ir}\int_{-\infty}^{\infty}\frac{dq}{(2\pi)^2}\frac{e^{iqr}}{q(1+\lambda^2q^2)}=\frac{1}{4\pi r}\left(1-e^{-r/\lambda}\right),
\end{equation}
for the Lorentz-invariant contribution, as derived in many textbooks.

In addition, we have
\begin{align}
B&=\frac{|\vec{v}|^2}{4\pi^2}\frac{1+3\cos(2\alpha)}{r^2}\int_\epsilon^\infty\frac{\cos(qr)dq}{q^4(1+\lambda^2q^2)^3} -\frac{|\vec{v}|^2}{4\pi^2 r^3}\int_\epsilon^\infty\frac{\sin(qr)dq}{q^5(1+\lambda^2q^2)^3}\nonumber\\&+ \frac{|\vec{v}|^2}{4\pi^2}\frac{(2\vartheta^2-1)}{r}\int_\epsilon^\infty\frac{\sin(qr)dq}{q^3(1+\lambda^2q^2)^3}+\frac{|\vec{v}|^2}{4\pi^2}\frac{\cos(2\alpha)}{r^3}\int_\epsilon^\infty\frac{(q^2r^2-3)\sin(qr)dq}{q^5(1+\lambda^2q^2)^3}.\label{eq08}
\end{align}
This set of important integrals that appears can be reduced to forms analogous to the following
\begin{align}
\int_\epsilon^\infty\frac{\cos{qr}dq}{q^4(1+\lambda^2q^2)^3}&\approx \int_\epsilon^\infty\left[\frac{1}{q^2}-3\lambda^2\right]\frac{\cos(qr)dq}{q^2}\nonumber\\&+\int_0^\infty\left[3+\frac{2}{(1+\lambda^2q^2)}+\frac{1}{(1+\lambda^2q^2)^2}\right]\frac{\lambda^4\cos(qr)dq}{(1+\lambda^2q^2)}\nonumber\\&=\frac{\pi}{48}\left[4 r \left(r^2+18\lambda^2\right)+3e^{-r/\lambda}\lambda\left(r^2+11r\lambda+35\lambda^2\right)\right]+\frac{2-3\epsilon^2(r^2+6\lambda^2)}{\epsilon^3},
\end{align}
such as,
\begin{align}
\int_\epsilon^\infty\frac{\sin{qr}dq}{q^5(1+\lambda^2q^2)^3}&\approx \int_\epsilon^\infty\left[\frac{1}{q^4}-\frac{3\lambda^2}{q^2}+6\lambda^4\right]\frac{\sin(qr)dq}{q}\nonumber\\&-\int_0^\infty\left[6+\frac{3}{(1+\lambda^2q^2)}+\frac{1}{(1+\lambda^2q^2)^2}\right]\frac{q\lambda^6\sin(qr)dq}{(1+\lambda^2q^2)}\nonumber\\&=\frac{\pi}{48}\left[r^4+36r^2\lambda^2+144\lambda^4-3e^{-r/\lambda}\lambda^2\left(r^2+13r\lambda+48\lambda^2\right)\right]\nonumber\\&+\frac{r}{6\epsilon^3}\left[2-\epsilon^2\left(r^2+18\lambda^2\right)\right].
\end{align}
Analogous procedures can be done for all of those integrals involving $\cos(qr)$ and $\sin(qr)$, which leads \eqref{eq08} to the following solution
\begin{align}
B&=\frac{|\vec{v}|^2}{6\pi^2\epsilon}(3\vartheta^2-2)+\frac{|\vec{v}|^2 e^{-\frac{r}{\lambda }}}{64 \pi  r^3} \Biggl\{r^4 \left[2 \vartheta ^2+\left(5-8 \vartheta ^2\right) e^{r/\lambda }-1\right]+2 \lambda 
   \left(9 \vartheta ^2-4\right) r^3\nonumber\\&-12 \lambda ^2 r^2 \left[-4 \vartheta ^2+\left(4 \vartheta ^2-3\right) e^{r/\lambda }+1\right]-48 \lambda ^4 \left(e^{r/\lambda }-1\right)+48 \lambda ^3 r\nonumber\\&+\cos (2
   \alpha ) \left[\left(12 \lambda ^2+r^2+6 \lambda  r\right)^2-e^{r/\lambda } \left(144 \lambda ^4+r^4-12 \lambda ^2
   r^2\right)\right]\Biggr\}\label{eq11}
\end{align}
in terms of the regulator parameter $\epsilon$. Adding $R_1(\epsilon)+R_2(\epsilon)$, one can see that the divergences present in both cancel out, leaving only a finite final result. The total contribution culminates in \eqref{eq02}, showing the complete expression for $V^{(1)}(r)$.

On the other hand, the second contribution of \eqref{eq01}, namely $V^{(2)}(r)$, has an analogous derivation, which, in addition to the previous calculation \eqref{eq11}, leads to the final result presented in \eqref{eq10}.}


\global\long\def\link#1#2{\href{http://eudml.org/#1}{#2}}
 \global\long\def\doi#1#2{\href{http://dx.doi.org/#1}{#2}}
 \global\long\def\arXiv#1#2{\href{http://arxiv.org/abs/#1}{arXiv:#1 [#2]}}
 \global\long\def\arXivOld#1{\href{http://arxiv.org/abs/#1}{arXiv:#1}}


\end{document}